\documentclass[12pt,fleqn]{revtex4}
\usepackage{latexsym}
\usepackage{amssymb}

\usepackage{amsfonts}

\def\J{\mathbf{J}}
\def\r{\mathbf{r}}
\def\v{\mathbf{v}}
\def\k{\mathbf{k}}

\def\u{\mathbf{u}}

\def\E{\mathbf{E}}
\def\B{\mathbf{B}}
\def\H{\mathbf{H}}
\def\D{\mathbf{D}}
\def\R{\mathbf{R}}

\def\F{\mathbf{F}}

\def\ext{\mathrm{ext}}
\def\eff{\mathrm{eff}}
\def\e{\mathrm{e}}

\def\rot{\mathrm{rot\,}}
\def\rdiv{\mathrm{div\,}}
\def\cth{\mathrm{cth}}
\def\Im{\mathrm{Im\,}}
\def\I{\mathrm{I}}
\def\L{\mathrm{L}}
\def\T{\mathrm{T}}
\def\Re{\mathrm{Re\,}}

\begin{document}

\title{To the Problem of Electromagnetic Field Energy in the Medium with
Temporal and Spatial Dispersion Outside the Transparency Domain}

\author{S.A. Trigger}
\email{satron@mail.ru}
\affiliation{Joint Institute for High Temperatures, Russian Academy of Sciences\\
Izhorskaya St. 13, Bd. 2, Moscow 125412, Russia}

\author{A.G. Zagorodny}
\email{azagorodny@bitp.kiev.ua}
\affiliation{Bogolyubov Institute for Theoretical Physics of the National Academy
of Sciences of Ukraine \\
14-b Metrolohichna St., Kiev 03680, Ukraine}

\begin{abstract}
The problem of calculation of electromagnetic field energy outside the transparency domain is discussed. It is shown that charged particle contribution to the energy of electromagnetic perturbations in the general case can be described in terms of bilinear combination of the dielectric polarizability of the medium. The explicit form of such contribution is found. The relations obtained are used to generalize the Planck's law to the case of absorptive medium.
\end{abstract}

\date{\today}

\maketitle

It is well known that the energy density of an electromagnetic wave in a
medium with spatial and temporal dispersion can be consistently defined only
in the transparency domain (see, for example, \cite{1,2,3} and references cited
therein). At the same time in the cited references there are no general
relations for the energy of electromagnetic field in the absorptive regions.

In spite of the fact that the idea of electromagnetic field energy
description in the general case was formulated many years ago \cite{4,5} and
some specific calculations for the medium with frequency dispersion outside
the transparency domain were made \cite{6,7} this problem requires further
consideration. The matter is that the energy of an electromagnetic
perturbation includes the ``pure'' electromagnetic energy and the kinetic
energy of charge carriers which they obtain due to their motion in the
electromagnetic field. If the neutral particles (i.e. atoms or molecules)
are present the additional potential energy acquired by bound electrons in
such field also should be added \cite{7}.

The purpose of the present contribution is to derive general relation for
the energy of electromagnetic perturbation in the medium with temporal and
spatial dispersion outside the transparency domain. We use the idea proposed
in \cite{4,5,6,7}, namely we treat the energy of perturbation as a sum of the
electromagnetic field energy and the particle energy (both kinetic and
potential) which particles acquire in the field. Obtained relations are
applied to calculate the fluctuation field energy and to generalize the
Planck formula for the case of non-transparent medium with spatial and
temporal dispersion.

We start from the Maxwell equations for electromagnetic field in a medium
in the form which is often uses in the plasma  theory \cite{2,3,8,9}
\begin{eqnarray}\label{eq1}
 \rot \E (\r,t) &=& - \frac{1}{c} \frac{\partial \B(\r,t)}{\partial t}, \nonumber\\
 \rdiv \B(\r,t) &=& 0, \nonumber\\
 \rot \B(\r,t) &=& \frac{1}{c} \frac{\partial \D(\r,t)}{\partial t} +
\frac{4\pi}{c} \J^e(\r,t), \nonumber\\
 \rdiv \D(\r,t) &=& 4\pi \rho^e(\r,t),
\end{eqnarray}
where $\J^e(\r,t)$ and $\rho^e(\r,t)$ are  the external sources, if present. In the case under consideration $\H(\r,t)\equiv\B(\r,t)$,  and total medium response to the electromagnetic field is described by dielectric permittivity tensor $\varepsilon_{ij}(\r,\r';t-t')$ or the conductivity tensor $\sigma_{ij}(\r,\r';t-t')$  \cite{8}
\begin{eqnarray}\label{eq2}
D_i(\r,t) &=& \int\limits_{-\infty}^t dt' \int d\r' \varepsilon_{ij}(\r,\r';t-t') E_j(\r',t') \nonumber\\
 J_i(\r,t) &=& \int\limits_{-\infty}^t dt' \int d\r' \sigma_{ij}(\r,\r';t-t') E_j(\r',t')
\end{eqnarray}
where $J_i(\r,t)$ is the total induced current which, by definition, includes all kinds of response. Therefore, the tensors  $\varepsilon_{ij}(\r,\r';t-t')$ and $\sigma_{ij}(\r,\r';t-t')$ are related as follows
\begin{equation}\label{eq3}
\varepsilon_{ij}(\r,\r';t-t') =\delta_{ij} \delta(\r-\r') \delta(t-t') +\int\limits_{t'}^t dt'' \sigma_{ij}(\r,\r';t''-t').
\end{equation}

We need also the equations describing interaction of electromagnetic fields with a medium. In what follows we illustrate the possibility to calculate the energy of electromagnetic perturbation outside the transparency domain using as an example of plasma-like medium. So, we supplement Eqs.~(1), (2) by the kinetic equation for plasma particles
\begin{eqnarray}\label{eq4}
\Bigl\{ \frac{\partial}{\partial t} +\v\frac{\partial}{\partial \r} + \frac{e_\alpha}{m_\alpha} \F^{\ext} +\frac{e_\alpha}{m_\alpha} \left[\E(\r,t) +\frac{\v}{c} \times \B(\r,t)\right] \frac{\partial}{\partial \v} \Bigr\}
\cdot f_\alpha(\r,\v,t) = \I_\alpha,
\end{eqnarray}
where $f_\alpha(\r,\v,t)$ is the distribution function of particles of $\alpha$ species, $\I_\alpha$ is the collision term, $\F^{\ext}$  is the external force field, if present, rest of notation is traditional.

Eq.~(\ref{eq4}) is valid in the case of classical plasma-like medium. The appropriate calculations for the case of combined plasma-molecular medium can be performed using the model of bound particles (see, for instance, Ref.~\cite{7,10,11}). Quantum description of both plasma and plasma-molecular systems is also possible \cite{9,11,12}. However, since the formulation of the general approach will not require the explicit form of the response function (except the calculation of specific examples) we need to know only general relation between the induced macroscopic currents $\J(\r,t)$ and self-consistent electric field $\E(\r,t)$ given by Eq.~(\ref{eq2}).

Using Eqs.~(\ref{eq1}) one obtains the well-known equation
\begin{equation}\label{eq5}
\frac{1}{4\pi} \left\{\E\frac{\partial \D}{\partial t} +\B\frac{\partial\B}{\partial t}\right\} +\J^{\ext} \E=-\frac{c}{4\pi} \rdiv[\E\B],
\end{equation}
which is reduced to the Pointing equation in the case of non-dispersive  medium. It can be used also to calculate the energy $W_\omega$ of quasi-monochromatic field in the case of weakly absorbing homogeneous medium \cite{7} and to recover
the well-known Brillouin formula
 \begin{equation}\label{eq6}
W_\omega=\frac{1}{16\pi} \left\{ \frac{\partial}{\partial\omega}
[\omega \varepsilon_{ij}(\omega)]
E_{\textbf{k},i}E^{*}_{\textbf{k},j}+B_{\textbf{k},i} B^{*}_{\textbf{k},j}\right\}.
\end{equation}
This equation is frequently used to calculate the energy density with regard to the influence of frequency dispersion \cite{13,14},
however Eq.~(\ref{eq6}) cannot be used in the case of strongly absorptive medium. In such case the treatment presented in Ref.~\cite{15} for lossless dissipative medium is also inappropriate. To get rid  of  the above mentioned restriction we derive equation for energy balance which takes into account the particle energy explicitly. This idea was suggested by V.~Ginzburg \cite{4,5}. In order to derive such equation it is necessary to multiply the kinetic equation (\ref{eq4}) by $n_\alpha m_\alpha v^2$  ($n_\alpha$ is the density of particle of $\alpha$ species) and integrate over the velocity $\v$. The result is
\begin{eqnarray}\label{eq7}
&&\frac{\partial}{\partial t}\int d \v\frac{n_\alpha m_\alpha  v^2}{2} f_\alpha(X,t) + \frac{\partial}{\partial\r} \int d \v\, \v \frac{n_\alpha m_\alpha  v^2}{2} f_\alpha(X,t) \nonumber\\
&+&\int d \v\frac{n_\alpha e_\alpha v^2}{2} \left[\E+\frac{v}{c}\times\B\right] \frac{\partial f_\alpha(X,t)}{\partial v}
= \int d\v \frac{m_\alpha v^2}{2} \I_\alpha.
\end{eqnarray}
Taking into account that $\int d\v \frac{m_\alpha v^2}{2} \I_\alpha=0$ and the equality
\begin{eqnarray}\label{eq8}
\sum_\alpha \int d \v\frac{n_\alpha e_\alpha v^2}{2}\left[\E+\frac{\v}{c}\times \B\right] \frac{\partial f_\alpha(X,t)}{\partial \v}
= -e_\alpha n_\alpha \int d \v\,\v \E f_\alpha(X,t)= -\E\J\ ,
\end{eqnarray}
and combining Eqs.~(\ref{eq7}), (\ref{eq8}) with the Eq.~(\ref{eq5}), which can be written in the form
\begin{eqnarray}\label{eq9}
\frac{1}{4\pi} \left\{\E\frac{\partial\E}{\partial t}+\B\frac{\partial\B}{\partial t}\right\} + \E\J +\J^e\E=-\frac{c}{4\pi}\rdiv[\E\B],
\end{eqnarray}
one obtains the equation for the energy balance \cite{5,6}
\begin{eqnarray}\label{eq10}
&&\frac{\partial}{\partial t}\Biggl\{ \frac{1}{8\pi} \left(\E^2(\r,t)+\B^2(\r,t)\right) + \sum_\alpha\int d \v
\frac{n_\alpha m_\alpha c v^2}{2} f_\alpha(X,t)\Biggr\}
 \nonumber\\
&+&\frac{\partial}{\partial \r}\left\{ \frac{c}{4\pi} \left[\E(\r,t)\times \B(\r,t)\right]+\sum_\alpha\int d \v\, \v
\frac{n_\alpha m_\alpha cv^2}{2} f_\alpha(X,t)\right\}  + \J^{\ext}(\r,t)\E(\r,t)=0,
\end{eqnarray}
where the terms responsible for the particle energy and energy flux are present in the explicit form. We see that there is no need to extract the particle energy term from the quantity $\E\frac{\partial\D}{\partial t}$ as it is done for derivation of Eq.~(\ref{eq6}). It is sufficient to use the solution of the kinetic equations.

Thus, the problem under consideration can be solved, if the distribution function is known. On the other hand, Eq.~(\ref{eq10}) makes it possible to use physical arguments to describe the particle energy  contribution to the energy of perturbation without restriction to the treatment of the case of transparent medium.

In the zero-order approximation on the gas-dynamic parameter ($l/L\ll1$, where $l$ is the mean free path, $L$ is the size of the system) the solution of the kinetic equation (\ref{eq4}) can be written in the form of the local Maxwellian distribution \cite{12}
\begin{equation}\label{eq11}
f_\alpha(X,t)=\frac{n_\alpha(\r,t)}{n_\alpha} \left(\frac{m_\alpha}{2\pi T_\alpha(\r,t)}\right)^{3/2} \exp\left[{-\frac{m_\alpha(\v-\u_\alpha(\r,t))}{2T_\alpha(\r,t)}}\right],
\end{equation}
where
\begin{eqnarray}\label{eq12}
n_\alpha(\r,t) &=& n_\alpha \int d\v f_\alpha(X,t),  \nonumber\\
u_\alpha(\r,t) &=& \frac{n_\alpha \int\v d\v f_\alpha(X,t)}{n_\alpha(\r,t)}, \nonumber\\
T_\alpha(\r,t) &=& \frac{n_\alpha \int d\v (m_\alpha/2)(\v-\u_\alpha(\r,t))^2f_\alpha(X,t)}{3 n_\alpha(\r,t)}.
\end{eqnarray}

Within such an approximation we can present the full $W$ energy density as follows
\begin{eqnarray}\label{eq13}
W &=& W_F+W_T+W_K =  \frac{1}{8\pi} \Bigl( \E^2(\r,t)+\B^2(\r,t)\Bigr) \nonumber\\
&+& \sum_\alpha \left( \frac{3}{2} n_\alpha(\r,t) T_\alpha(\r,t) + \frac{m_\alpha n_\alpha(\r,t) \u_\alpha^2(\r,t)}{2} \right),
\end{eqnarray}
where the field $W_F$, thermal $W_T$ and kinetic $W_K$  energies are equal respectively
\begin{eqnarray}\label{eq14}
 W_F &=& \frac{1}{8\pi} \left( \E^2(\r,t)+\B^2(\r,t)\right), \nonumber\\
W_T &=& \sum_\alpha \frac{3}{2} n_\alpha(\r,t) T_\alpha(\r,t), \nonumber\\
W_K &=& \sum_\alpha n_\alpha(\r,t) \frac{m_\alpha u_\alpha^2(\r,t)}{2}.
\end{eqnarray}
Since $W_T$ is the heat produced by perturbation we can treat the energy associated with the electromagnetic field as the sum of $W_F$ and $W_K$.

Restricting ourselves by the second order approximation in perturbation, we can rewrite the part of energy $W_K$ as
\begin{equation}\label{eq15}
 W_K =\sum_\alpha \frac{n_\alpha m_\alpha \u_\alpha^2(\r,t)}{2} = \sum_\alpha \frac{m_\alpha}{2e_\alpha^2 n_\alpha}\, J_\alpha^2(\r,t).
\end{equation}
Here $J_\alpha(\r,t)$ is the partial contribution of the particle of $\alpha$ species to the induced current \linebreak $J(\r,t)=\sum_\alpha J_\alpha(\r,t)$.

Notice, that Eq.~(\ref{eq15}) directly follows from the transparent physical treatment: the kinetic energy acquired by particles under the action of electromagnetic field can be directly expressed in terms of the averaged induced velocity. Namely this approach was used by V.~Ginzburg to estimate the energy density of particles in the case of cold plasmas. However, as is seen Eq.~(\ref{eq15}) do not require such restriction.

The generalization of the results obtained in \cite{6,7} can be achieved using the relation between the induced current and electric field (\ref{eq2}) that gives
\begin{eqnarray}\label{eq16}
W_K &=& \sum_\alpha \frac{m_\alpha}{2e_\alpha^2 n_\alpha}\int\limits_{-\infty}^t dt' \int d\r' \sigma_{ij}^{(\alpha)}(\r,\r';t,t')
\cdot\int\limits_{-\infty}^t dt'' \int d\r'' \sigma_{ij}^{(\alpha)}(\r,\r'';t,t'') E_j(\r',t')E_\k(\r'',t'') \nonumber\\
&=&\sum_\alpha \frac{m_\alpha}{2e_\alpha^2 n_\alpha} \int \frac{d\omega}{2\pi} \int \frac{d\k}{(2\pi)^3} \int \frac{d\omega}{2\pi} \int \frac{d\k'}{(2\pi)^3} \nonumber\\
&\cdot& \e^{-i(\omega-\omega')t} \e^{i(\k-\k')\r} \sigma_{ij}^{(\alpha)}(\k,\omega) \sigma_{ik}^{(\alpha)*}(\k',\omega') E_{i\k\omega}E_{j\k'\omega'}^*\, ,
\end{eqnarray}
where $\sigma_{ij}^{(\alpha)}(\k,\omega)$ is the partial contribution of particles of $\alpha$ species to the conductivity tensor of the system 
\[
\sigma_{ij}^{(\alpha)}(\k,\omega) = \sum_\alpha \sigma_{ij}^{(\alpha)}(\k,\omega),
\]
or in terms of the generalized polarizability $\chi_{ij}^{(\alpha)}(\k,\omega)\equiv\frac{4\pi i}{\omega} \sigma_{ij}^{(\alpha)}(\k, \omega)$
the effective energy of electromagnetic perturbation in the medium $W_F^{\eff}\equiv W_F+W_K$
can be written as
\begin{eqnarray}\label{eq17}
W_F^{\eff} &=& \frac{1}{8\pi} \int\frac{d\k}{(2\pi)^3}\int\frac{d\k'}{(2\pi)^3} \int\frac{d\omega}{2\pi}\int\frac{d\omega'}{2\pi}\, \e^{i(\k-\k')\r} \e^{-i(\omega-\omega')t} \nonumber\\
&\cdot&  \Biggl\{ \frac{k_ik'_j}{\k\k'} +\left(1+\frac{c^2}{\omega\omega'} \k\k'\right) \left(\delta_{ij}-\frac{k_i k'_j}{\k\k'}\right)
+\sum_{\alpha=e,i} \frac{\omega^2}{\omega_{p\alpha}^2} \chi_{ki}^{(\alpha)}(\k,\omega)\chi_{\k j}^{(\alpha)*}(\k',\omega')\Biggr\} E_{i\k\omega} E_{j\k'\omega'}^*\, ,
\end{eqnarray}
where $\omega_{p\alpha}^2=4\pi e_\alpha^2 n_\alpha/m_\alpha$.

This is the general relation for the electromagnetic perturbation energy in plasma-like medium.

Notice, that Eq.~(\ref{eq15}) can be used also to estimate the kinetic energy of  bound electrons in atoms and molecules. However, in such a case along with the kinetic energy of electrons the energy of electromagnetic perturbation includes the potential energy of bound electrons in the fields of ions with which they are bound. In the case of classical model of atom-oscillator \cite{7,12} such energy can be estimated as
\[
W_{U} = n_m \frac{\omega_0^2 r_m^2(\r,t)}{2}.
\]
Here $n_m$ is the density of bound electrons, $\omega_0$ is the eigenfrequency of the oscillator, $\r_m(\r,t)$ is the  oscillation amplitude of the bound electron. Since
$\u_m(\r,t)=\frac{d\r_m(\r,t)}{dt}$,
the energy $W_B$ can be expressed in terms of the mean velocity $\u_m(\r,t)$, i.e. in terms of the induced current of the bound electrons. Thus,
\begin{eqnarray}\label{18}
W_{U} &=& \frac{1}{8\pi} \int \frac{d\k}{(2\pi)^3} \int\frac{d\k'}{(2\pi)^3} \int\frac{d\omega}{2\pi} \int\frac{d\omega'}{2\pi} \, \e^{i(\k-\k')\r}\nonumber\\
&\cdot&  \e^{-i(\omega-\omega')t} \frac{\omega_0^2}{\omega_{pm}^2} \chi_{ki}^{(m)}(\k,\omega) \chi_{lj}^{(m)*}(\k',\omega') E_{i\k\omega} E_{j\k'\omega'}^*,
\end{eqnarray}
where $\chi_{ij}^{(m)}(\k,\omega)$ in the case of classical model of atom-oscillator is given by \cite{10}
\begin{equation}\label{eq19}
\chi_{ij}^{(m)}(\k,\omega) =-\delta_{ij}\int d\v \frac{\omega_{pm}^2 f_{0m}(\v)}{(\omega-\k\v)^2-\omega_0^2+i\gamma(\omega-\k\v)},\qquad \omega_{pm}^2=\frac{4\pi e_b^2 n_m}{m_b},
\end{equation}
$f_{0m}(\v)$ is the distribution function of bound particles (atoms, or molecules), $e_b$ and $m_b$ are the effective charge and the reduced mass of bound electron.

So, in the case of plasma-molecular system the energy of perturbation can be written as
\begin{eqnarray}\label{eq20}
W &=& W_F^{\eff}+W_{U}= W_F+W_K+W_{U}
= \frac{1}{8\pi} \int\frac{d\k}{(2\pi)^3} \int\frac{d\k'}{(2\pi)^3} \int\frac{d\omega}{2\pi} \int\frac{d\omega'}{2\pi} \, \e^{i(\k-\k')\r} \;\e^{-i(\omega-\omega')t}  \nonumber\\
&\cdot& \Biggl\{ \left(\delta_{ij}-\frac{k_i k'_j}{\k\k'}\right) \left(1+\frac{c^2\k\k'}{\omega^2}\right) +\frac{k_i k'_j}{\k\k'}
+\sum_{\alpha=e,i} \frac{\omega^2}{\omega_{p\alpha}^2} \chi_{ki}^{(\alpha)}(\k,\omega) \chi_{kj}^{(\alpha)*}(\k',\omega')\nonumber\\
&+&  \frac{\omega^2+\omega_0^2}{\omega_{pm}^2} \, \chi_{ki}^{(m)}(\k,\omega) \chi_{kj}^{(m)*}(\k',\omega') \Biggr\}
\cdot E_{i\k\omega} E_{j\k'\omega}^*.
\end{eqnarray}

Notice, that this equation remains valid in the case of quantum description, if the polarizabilities  $\chi_{ij}^{(\alpha)}(\k,\omega)$ ($\alpha=e,i,m$) are calculated within the framework of quantum theory (see, for example, Ref.~\cite{11,12}).

In the case of the monochromatic field
$\E(\r,t)=\frac{1}{2} \left\{ \E(\r)\e^{-i\omega t} +\E^*(\r) \e^{i\omega t}\right\}$,
after the averaging over the period of oscillation $T=2\pi$ and the volume of the system $V$, Eq.~(\ref{eq20}) is reduced to
\begin{eqnarray}\label{eq21}
\overline{W} &\equiv& \overline{W}_F^{\eff}+\overline{W}_{U}= \frac{1}{16\pi V} \int\frac{d\k}{(2\pi)^3} \Biggl\{ \left(\delta_{ij}-\frac{k_i k_j}{k^2}\right)\left(1+\frac{c^2k^2}{\omega^2}\right) +\frac{k_ik_j}{k^2} \nonumber\\
&+&\sum_{\alpha=e,i} \frac{\omega^2}{\omega_{p\alpha}^2} \, \chi_{ki}^{(\alpha)}(\k,\omega) \chi_{kj}^{(\alpha)*}(\k,\omega)
+ \frac{\omega^2+\omega_0^2}{\omega_{pm}^2}\, \chi_{ki}^{(m)}(\k,\omega) \chi_{kj}^{(m)*}(\k,\omega)\Biggr\} \overline{E_{\k i}E_{\k j}^*}.
\end{eqnarray}

Let us consider the case when the spatial dispersion is formally neglected, i.g., we put
$\chi_{ij}^{(\alpha)}(\k,\omega) = \chi_{ij}^{(\alpha)}(\omega)$. In this case, equation (\ref{eq21}) is simplified to
\begin{equation}\label{eq22}
\overline{W}\equiv\overline{W}_E+\overline{W}_B,
\end{equation}
where
\begin{eqnarray}\label{eq23}
\overline{W}_E &=& \frac{1}{16\pi} \Biggl\{ \delta_{ij}+\sum_{e,i,m} \frac{\omega^2+\omega_{0\alpha}^2}{\omega_{p\alpha}^2}\,
\chi_{ki}^{(\alpha)}(\omega) \chi_{kj}^{(\alpha)*}(\omega)\Biggr\}
\overline{E_i E^*_j},\nonumber\\
\overline{W}_B &=& \frac{1}{16\pi}\overline{|\B|^2} ,\quad \overline{|\B|^2} =\frac{1}{V} \int d\r |\B(\r)|^2, \nonumber\\
\overline{E_i E^*_j} &=& \frac{1}{V} \int d\r E_i(\r) E_j^*(\r), \qquad \qquad\qquad
\omega_{0\alpha} = \left\{ \begin{array}{cc}
0, & \alpha=e,i\\
\omega_0, & \alpha=m \end{array} \right.\, .
\end{eqnarray}

Using (\ref{eq22}), (\ref{eq23}) it is easy to recover the results obtained in Refs.~\cite{6,7} for the electric field energy density outside the transparency domain. For example, in the case of cold molecular system
\begin{equation}\label{eq24}
\chi_{ij}^{(m)}(\omega) =-\delta_{ij}\frac{\omega_{pm}^2}{\omega^2-\omega_0^2+i\gamma\omega},
\end{equation}
that leads to
\begin{equation}\label{eq25}
\overline{W}_E =\frac{1}{16\pi} \overline{|\E|^2} \left[ 1+\frac{\omega_{pm}^2 (\omega^2+\omega_0^2)}{(\omega^2-\omega_0^2)^2+\gamma^2\omega^2}\right].
\end{equation}

In the case of cold plasma
\begin{equation}\label{eq26}
\chi_{ij}^{(e)}(\omega) =-\delta_{ij}\frac{\omega_{pe}^2}{\omega(\omega+i\nu_e)},
\end{equation}
where $\nu_e$ is the effective collision frequency, that gives
\begin{equation}\label{eq27}
\overline{W}_E=\frac{1}{16\pi} \left[ 1+\frac{\omega_{pe}^2}{\omega^2+\nu_e}\right] \overline{|\E|^2}.
\end{equation}
Eqs.~(\ref{eq25}) and (\ref{eq27}) are in agreement with the well-known relation (see the first term in  Eq.~(\ref{eq6}))
\begin{equation}\label{eq28}
\overline{W} =\frac{1}{16\pi} \frac{\partial}{\partial\omega} \left[\omega \Re\varepsilon_{ij}(\omega)\right] \overline{E_i E_j^*}
\end{equation}
only in the case of nondissipative systems ($\gamma=0$ and $\nu=0$).

It is easy to show that Eq.~(\ref{eq21}) can be also applied to description of the energy density  of fluctuation. Performing statistical averaging of Eq.~(\ref{eq21}), one obtains
\begin{eqnarray}\label{eq29}
\langle W\rangle &=& \frac{1}{8\pi} \int\frac{d\k}{(2\pi)^3} \int\frac{d\omega}{2\pi} \Biggl\{ \frac{k_i k_j}{k^2} +
\left(1+\frac{c^2k^2}{\omega^2}\right)
\left( \delta_{ij}-\frac{k_i k_j}{k^2}\right) \nonumber\\
&+& \sum_{\alpha=e,i,m} \frac{\omega^2+\omega_{0\alpha}^2}{\omega_{p\alpha}^2}\, \chi_{ki}^{(\alpha)}(\k,\omega) \chi_{kj}^{(\alpha)*}(\k,\omega) \Biggr\} \langle \delta E_i \delta E_j\rangle_{\k\omega}.
\end{eqnarray}
Deriving Eq.~(\ref{eq29}) we take into account that
\begin{equation}\label{eq30}
\langle\delta E_{i\k\omega} \delta E_{j\k'\omega'}^*\rangle = (2\pi)^4 \delta(\k-\k') \delta(\omega-\omega') \langle \delta E_i \delta E_j\rangle_{\k\omega},
\end{equation}
where
\begin{eqnarray}\label{eq31}
\langle \delta E_i \delta E_j\rangle_{\k\omega}= \int \! d\R \e^{-i\k\R}\! \int d\omega \e^{+i\omega\tau} \langle \delta E_i(\r,t) \delta E_j(\r',t')\rangle_{\k\omega},\qquad
\R = \r-\r', \qquad \tau=t-t'.
\end{eqnarray}

In the case of equilibrium system $\langle \delta E_i \delta E_j\rangle_{\k\omega}$ is given by the fluctuation dissipation theorem (see, for example, \cite{2,3})
\begin{equation}\label{eq32}
\langle \delta E_i \delta E_j\rangle_{\k\omega} =\frac{4\pi i}{\omega} \theta(\omega) \left\{\Lambda_{ij}^{-1}(\k,\omega)-\Lambda_{ji}^{-1*}(\k,\omega)\right\}.
\end{equation}
Here
\begin{eqnarray}\label{eq33}
\theta\equiv \frac{\hbar\omega}{2}\cth \frac{\hbar\omega}{2T}, \qquad
\Lambda_{ij}(\k,\omega) = \varepsilon_{ij}(\k,\omega) -\frac{k^2c^2}{\omega^2}\left(\delta_{ij}-\frac{k_ik_j}{k^2}\right).
\end{eqnarray}

Further simplification of (\ref{eq29}), (\ref{eq33}) can be done in the case of isotopic system for which
\begin{equation}\label{eq34}
\varepsilon_{ij}(k,\omega) =\varepsilon_\T(\k,\omega) \left(\delta_{ij}-\frac{k_ik_j}{k^2}\right) +\varepsilon_\L(\k,\omega) \frac{k_ik_j}{k^2},
\end{equation}
where $\varepsilon_\T(\k,\omega)$ and $\varepsilon_\L(\k,\omega)$ are the transverse and longitudinal parts of the dielectric permittivity tensor.

Substitution of (\ref{eq34}) into (\ref{eq32}) and (\ref{eq29}) yields
\begin{equation}\label{eq35}
\langle W\rangle =\int\limits_0^\infty \langle W\rangle_\omega d\omega,
\end{equation}
where for the general case of the non-transparent medium
\begin{eqnarray}\label{eq36}
\langle W\rangle_\omega &=& \frac{\theta(\omega)}{2\pi^3\omega}\int\limits_0^\infty dk\, k^2 \Biggl\{ \frac{\Im\varepsilon_\L(k,\omega)}{|\varepsilon_\L(k,\omega)|^2}
\cdot\Biggl[1+ \sum_{e,i,m}\frac{\omega^2+\omega_{0\alpha}^2}{\omega_{p\alpha}^2} |\chi_\L^{(\alpha)}(\k,\omega)|^2\Biggr] \nonumber\\
&+& \frac{2\Im\varepsilon_\T(k,\omega)}{|\varepsilon_\T(k,\omega)-(k^2c^2)/(\omega^2)|^2}
\cdot\Biggl[ 1+ \frac{k^2c^2}{\omega^2} +\sum_{e,i,m} \frac{\omega^2+\omega_{0\alpha}^2}{\omega_{p\alpha}^2} |\chi_\T^{(\alpha)}(\k,\omega)|^2\Biggr]\Biggr\}.
\end{eqnarray}

In the case of negligible dissipation we can use the approximation of the type
\begin{eqnarray}\label{eq37}
\frac{\Im \varepsilon_\T}{|\varepsilon_\T(k,\omega)-(k^2c^2)/(\omega^2)|^2}  \simeq  \pi\delta \left(\Re\varepsilon_\T(k,\omega)-\frac{k^2c^2}{\omega^2}\right).
\end{eqnarray}

In the case of cold plasma at $\omega\gg\nu$
\[
\varepsilon_\T(\omega)\simeq 1-\frac{\omega_p^2}{\omega^2}
\]
and
\begin{eqnarray}\label{eq38}
\langle W\rangle_\omega \!=\! \frac{\omega^2\theta(\omega)}{2\pi^2 c^3}\sqrt{\varepsilon(\omega)} \left(1\!+\varepsilon(\omega)\!+\frac{\omega_{pe}^2}{\omega^2}\right)\! =\!
 \frac{\omega^2\theta(\omega)}{\pi^2 c^3}\sqrt{\varepsilon(\omega)}
\end{eqnarray}
This result is in agreement with the well-known result for the energy density in the dispersive  transparent medium \cite{16}.
If we rewrite (\ref{eq38}) in the form
\begin{eqnarray}\label{eq39}
\langle W\rangle_\omega =
 \frac{\hbar\omega^3}{\pi^2c^3}\left\{\frac{1}{2}+\frac{1}{\e^{(\hbar\omega)/T}-1}\right\}\sqrt{\varepsilon(\omega)}
\end{eqnarray}
we can interpret the second term in the sum (\ref{eq39}) as the generalization of the Planck formula for the transparent plasma. This generalization follows from straightforward consideration of the transverse oscillations $\omega=\sqrt{c^2k^2+\omega_p^2}$ in plasma as a non-damping boson quasiparticles \cite{17}.

In the case of molecular medium $\gamma\to0$
\begin{equation}\label{eq40}
\varepsilon_b(\omega) =1-\frac{\omega_{pb}^2}{\omega^2-\omega_0^2}
\end{equation}
and thus
\[
\langle W\rangle_\omega = \frac{\omega^2\theta(\omega)}{2\pi^2 c^3}\sqrt{\varepsilon_b(\omega)} \left[1+\varepsilon_b(\omega)+\frac{\omega_{pm}^2(\omega^2+\omega_0^2)}{(\omega^2-\omega_0^2)^2}\right]\, .
\]
At $\omega\gg\omega_0$ we come back to the equation of the type (\ref{eq38})
\begin{equation}\label{eq41}
\langle W\rangle_\omega = \frac{\omega^2\theta(\omega)}{\pi^2 c^3}\sqrt{\varepsilon_b(\omega)} .
\end{equation}
At $\omega\ll\omega_0$ the frequency dispersion can be neglected and we obtain the result for nondispersive transparent medium \cite{16}
\begin{equation}\label{eq42}
\langle W\rangle_\omega = \frac{\omega^2\theta(\omega)}{2\pi^2 c^3}\tilde{\varepsilon}^{3/2} ,
\end{equation}
where
\[
\tilde{\varepsilon}=\lim\limits_{\omega\to 0}\, \varepsilon_b(\omega).
\]

 In the general case Eq.~(\ref{eq36}) can be rewritten in the form of the Planck formula modified by the presence of the medium
\begin{equation}\label{eq43}
\langle W\rangle_\omega = \frac{\hbar\omega^3}{\pi^2 c^3}\left\{ \frac{1}{2}+\frac{1}{\e^{(\hbar\omega)/T}-1}\right\} S(\omega),
\end{equation}
where $S(\omega)$ is the function describing the influence of the medium
\begin{eqnarray}\label{eq44}
S(\omega) &=& \frac{c^3}{2\pi\omega^3} \int\limits_0^\infty dk\, k^2 \Biggl\{ \frac{\Im \varepsilon_\L(k,\omega)}{|\varepsilon_\L(k,\omega)|^2}
\cdot\left[ 1+\sum_{e,i,m}\frac{\omega^2+\omega^2_{0\alpha}}{\omega^2_{p\alpha}} |\chi_\L^{(\alpha)}(\k,\omega)|^2\right]\nonumber\\ &+& \frac{2\Im \varepsilon_\T(k,\omega)}{|\varepsilon_\T(k,\omega)-\frac{k^2c^2}{\omega^2}|^2}
\cdot \Biggl[ 1+\frac{k^2c^2}{\omega^2}
+ \sum_{e,i,m}\frac{\omega^2+\omega^2_{0\alpha}}{\omega^2_{p\alpha}} |\chi_\T^{(\alpha)}(\k,\omega)|^2\Biggr]\Biggr\}.
\end{eqnarray}

Notice, that the alternative  approach on the basis of quantized electromagnetic field has been recently proposed in Ref.~\cite{18}. The detailed comparison on these two approaches will be the matter of the separate treatment. 

Thus, in the present contribution we derive the general relations for the energy density of electromagnetic perturbation in absorptive medium.
This approach is based on the equation for the energy balance in the medium ((\ref{eq10})) and the assumption, that the distribution function for particles subordinates to the local Maxwellian distribution (\ref{eq11}).
The representation (\ref{eq43}) describes the  generalized zero oscillations, depending on temperature and density of medium via the function $S(\omega)$ (the first term in the brackets) and the modification of the Planck formula (the second term in the brackets).

This work was supported by the Target Program of the Physics and Astronomy Department of the National Academy of Sciences of Ukraine, No.~0117U000240.


\begin{thebibliography}{99}

\bibitem{1} L.D.~Landau, E.M.~Lifshits. \textit{Electrodynamics of Continuous Medium}. Pergamon, 1960, 413 p.

\bibitem{2} A.I.~Akhiezer, I.A.~Akhiezer, A.G.~Sitenko, K.M.~Stepanov, R.V.~Polovin. \textit{Plasma Electrodynamics. Volume 1. Linear Theory}. N.-Y.: Pergamon, 1975, 431 p.

\bibitem{3} A.G.~Sitenko, V.M.~Malnev. \textit{Plasma Physics Theory}. Chapman and Hall, 1994, 432 p.

\bibitem{4} V.L.~Ginzburg. Radiofizika. Izv. Vuzov, \textbf{4}, 74 (1961).

\bibitem{5} B.N.~Gershman, V.L.~Ginzburg. Radiofizika. Izv. Vuzov,  \textbf{5}, 31 (1962).

\bibitem{6} V.L.~Ginzburg. \textit{The Propagation of Electromegnetic Waves in Plasmas}. London:~Pergamon Press, 1964, 535 p.

\bibitem{7} Yu.S.~Barash, V.L.~Ginzburg. Sov. Phys. Uspekhi, \textbf{19}, 263-270 (1976).

\bibitem{8} S.~Ichimaru. \textit{Basic Principles of Plasma Physics. A Statistical Approach}. Reading, Mass.: Benjamin, 1973, 324~p.

\bibitem{9} V.P.~Silin, A.A.~Rukhadze. \textit{Electromagnetic Properties of Plasmas and Plasma-Like Media}. Moscow: Gosatomizdat, 1961.

\bibitem{10}Yu.L.~Klimontovich, H.~Wilhelmsson, I.P.~Yakimenko, A.G.~Zagorodny. Phys.Rep., \textbf{175,} 263 (1989).

\bibitem{11} Yu.L. Klimontovich, AY Shevchenko, IP Yakimenko, AG Zagorodny.
Contributions to Plasma Physics, \textbf{29} 551 (1989).

\bibitem{12} Yu.L.~Klimontovich. \textit{Statistical Physics}. N.Y.: Harwood, 1986, 734~p.

\bibitem{13} F.S.S.~Rosa, D.A.R.~Dalvit, and P.W.~Miloni. Phys. Rev. A \textbf{81}:3, 033812 (2010).

\bibitem{14} J.M.~Zhao, Z.M.~Zhang. J. Quant. Spectrosc. Radiat. Transfer, \textbf{151}, 49-57 (2015).

\bibitem{15} T.G.~Philbin. Phys. Rev. A, \textbf{83}, 013823 (2011).


\bibitem{16} M.L.~Levin, S.M.~Rytov. \textit{The Theory of Equilibrium Thermal Fluctuations in Electrodynamics}. M.:~Nauka, 1967, 308~p.

\bibitem{17} S.A. Trigger. Phys. Lett. A, \textbf{370}, 365 (2007).


\bibitem{18}V.B. Bobrov, S.A. Trigger. Theor. Math. Phys., \textbf{187}, 520 (2016).


\end{thebibliography}
\end{document}